\documentclass{aipproc}
\def\selectedlayoutstyle{6x9}
\layoutstyle\selectedlayoutstyle
\usepackage{epsfig}
\SetInternalRegister\hbadness{8000} 

\newcommand\doingARLO[2][]{%
  \ifx\mmref\undefined #1\else #2\fi
}

\begin{document}

\title
      []
{CLEO-c and CESR-c: A New Frontier in Weak and Strong
Interactions\\}


\author{Ian Shipsey}{
  address={1396 Physics Building, West Lafayette, IN 47907, U.S.A.},
  email={shipsey@physics.purdue.edu},
  thanks={}
}


%
%

%

\copyrightyear  {2001}

\begin{abstract}
We report on the physics potential of a proposed conversion of the
CESR machine and the CLEO detector to a charm and QCD factory:
``CLEO-c and CESR-c'' that will make crucial contributions to
flavor physics in this decade and offers our best hope for
mastering non-perturbative QCD which is essential if we are to
understand strongly coupled sectors in the new physics that lies
beyond the Standard Model.
\end{abstract}

\date{\today}

\maketitle

\section{Executive Summary}

A focused three year program of charm and QCD physics with the CLEO
detector operating in the 3-5 GeV energy range at the Cornell Electron Storage
Ring in Ithaca, NY, is proposed.  The CLEO-c physics program includes
a set of measurements that will substantially advance our understanding
of important processes within the Standard Model of particle physics
and set the stage for understanding the larger theory in which we imagine the
Standard Model to be embedded.  The program will be preceded by
one year of bottomium running with CLEOIII in 2002.

The program revolves around the strong interaction and the
pressing need to develop sufficiently powerful tools to deal with
an intrinsically non-perturbative theory. At the present time, and
for the last twenty years, progress in weak interaction physics
and the study of heavy flavor physics has been achieved primarily
by seeking the few probes of weak-scale physics that succesfully
evade or minimise the role of strong interaction physics. The
pre-eminence of the mode $B \rightarrow J/\psi K^o_S$ in measuring
$sin 2 \beta$ arises from the absence of complications due to the
strong interaction. Similarly, Heavy Quark Symmetry and the
development of HQET enabled the identification of the zero-recoil
limit as the best place to measure $b \rightarrow c \ell \nu$
exclusive decays as the strong intercation effects at this
kinematic point are small and calculable. This method has become
one of the two preferred ways to extract $V_{cb}$. If we had
similar strategies that would allow us to extract $V_{ub}$ from $b
\rightarrow u \ell \nu$ without form factor uncertainties, and
$V_{td}$  and $V_{ts}$ from  $B_d$ and $B_s$ mixing respectively,
without decay constant and bag parameter uncertainties, we would
be well on the way to understanding the CKM matrix at the few per
cent level.

The goal of flavor physics is to test the Standard Model description
of CP violation by probing the unitarity of the CKM matrix.
However this goal is in jeopardy. Across the spectrum of heavy quark
physics the study of weak-scale phenomena and the extractiion of quark
mixing paramters remain fundamentally limited by our
restricted capacity to deal with the non-perturbative strong interaction dynamics.

\begin{figure}
\epsfig{figure=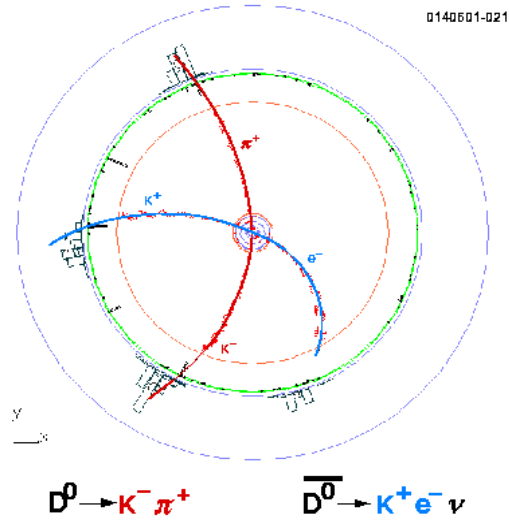,width=7cm,height=7cm}
\caption {A doubly tagged  event at the  $\psi(3770)$}
\label{fig:cleoc_event}
\end{figure}

Moreover, as we look to the future beyond the
Standard Model, we anticipate that the larger
theory in which the Standard Model is embedded will likely
be strongly coupled, or have strongly coupled sectors.
Both Technicolor, which is modeled on QCD and
is ab initio strongly coupled, and Supersymmetry,
which employs strongly coupled sectors to break the supersymmetry,
are prime examples of candidates for physics beyond the Standard Model.
Strong coupling is a phenomenon to be expected, weak coupling is the
exception in field theory not the norm. However our ability to
compute reliably in a strongly coupled theory is not well developed. Techniques
such as lattice gauge theory that deal
squarely with strongly coupled theories will eventually
determine our progress on all fronts in particle physics. At the
present time the absence of adequate theoretical tools significantly
limits the physics we can obtain from the study heavy quarks.

Recent advances in Lattce QCD (LQCD) may offer hope.
Algorithmic advances and  improved computer hardware have
produced a wide variety of non-perturbative
results with accuracies at the 10-20\%
level for systems involving one or two heavy
quark such as  $B$ and $D$  mesons, and $\Psi$ and $\Upsilon$
quarkonia. First generation unquenched calculations
have been completed for decay constants and
semileptonic form factors, for mixing and spectra. There is very strong
interest in the LQCD community in pursuing much higher precision
and the techniques needed to reduce uncertainties to 1-2\% precision. But the push
towards higher precision is hampered by the absence of accurate charm
data against which to test and calibrate the new theoretical techniques.

CLEO-c will address this challenge in the charm system at threshold where the experimental
conditions are optimal.   We will obtain data samples
several hundred times larger than any previous experiment
and with a detector which is an order of magnitude better
than any previous detector to operate at charm threshold.
We will supplement the charm and charmonium
data with bottomium data taken starting Fall 2001 with CLEO III prior
to the conversion to CLEO-c. Decay constants, form factors,
spectroscopy of open and hidden charm and hidden bottom,
and an immense variety of absolute branching ratio measurements
will be obtained with accuracies at the 1-2\% level. Precision measurements will demand precision theory.

The measurements described here are an essential and integral part
of the global program in heavy flavor physics of this decade,
and the larger program of the as yet unknown physics of the next decade.

\section{Introduction}

\begin{figure}
\epsfig{figure=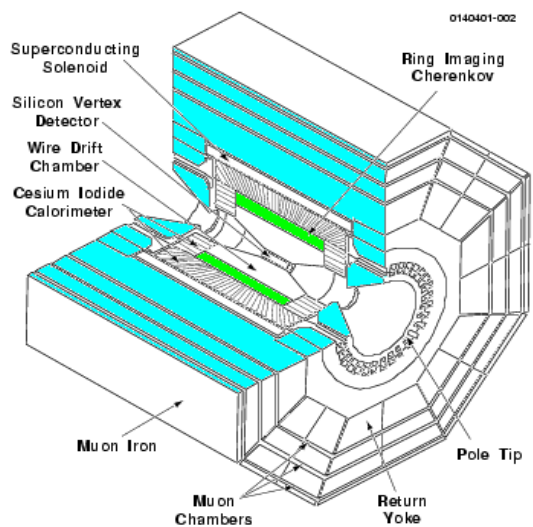,width=7cm,height=7cm}
\caption {The CLEO III detector.}
\label{fig:cleo3_det}
\end{figure}

For many years, the CLEO experiment at the Cornell Electron Storage
Ring, CESR, operating on the $\Upsilon$(4S) resonance, has provided most of the
world's information about the $B_{d}$ and $B_{u}$ mesons.
At the same time, CLEO, using the copious continuum pair production at the
$\Upsilon$(4S) resonance
has been a leader in the study of charm and $\tau$ physics. Now that the
asymmetric B factories have achieved high luminosity, CLEO is
uniquely positioned  to advance
the knowledge of heavy flavor physics by carrying out several measurements
near charm threshold, at center of mass energies in the 3.5-5.0 GeV region.
These measurements address crucial topics which benefit from the high
luminosity and experimental constraints which exist near threshold but
have not been carried out at existing charm factories because the luminosity
has been too low, or have been carried out previously with meager statistics.
They include:
\begin{enumerate}
\item Charm Decay constants $f_D, f_{D_s}$
\item Charm Absolute Branching Fractions
\item Semileptonic decay form factors
\item Direct determination of $V_{cd}$ \& $V_{cs}$
\item QCD studies including: \\
Charmonium and bottomonium spectroscopy \\
Glueball and exotic searches \\
Measurement of R between 3 and 5 GeV, via scans \\
Measurement of R between 1 and 3 GeV, via ISR
\item Search for new physics via charm mixing, CP violation
and rare decays
\item $\tau$ decay physics
\end{enumerate}

The CLEO detector can carry out this program with only minimal
modifications. The CLEO-c project is described at length in
\cite{cleo-c} - \cite{maravin}. A very modest upgrade to the
storage ring is required to achieve the required luminosity.
Below, we summarize the advantages of running at charm threshold,
the minor modifications required to optimize the detector,
examples of key analyses, a description of the proposed run plan,
and a summary of the physics impact of the program.

\begin{table}[t]
\caption{Summary of CLEO-c charm decay measurements.}
\label{tab:charm}
\begin{tabular}{cccccc}
\hline
Topic & Reaction & Energy & $L $        & current     & CLEO-c \\
      &          & (MeV)  & $(fb^{-1})$ & sensitivity & sensitivity \\
\hline
\multicolumn{1}{c}{Decay constant} &
\multicolumn{5}{c}{}\\
\hline
$f_D$ & $D^+\to\mu^+\nu$ & 3770 & 3 & UL & 2.3\% \\
\hline
$f_{D_s}$ & $D_{s}^+\to\mu^+\nu$ & 4140 & 3 & 14\% & 1.9\% \\
\hline
$f_{D_s}$ & $D_{s}^+\to\mu^+\nu$ & 4140 & 3 & 33\% & 1.6\% \\
\hline
\multicolumn{2}{c}{Absolute Branching Fractions} &
\multicolumn{4}{c}{}\\
\hline
\multicolumn{2}{c}{$Br(D^0 \to K\pi)$} & 3770 & 3 & 2.4\% & 0.6\% \\
\hline
\multicolumn{2}{c}{$Br(D^+ \to K\pi\pi)$} & 3770 & 3 & 7.2\% & 0.7\% \\
\hline
\multicolumn{2}{c}{$Br(D_s^+\to\phi\pi)$} & 4140 & 3 & 25\% & 1.9\% \\
\hline
\multicolumn{2}{c}{$Br(\Lambda_c\to pK\pi)$} & 4600 & 1 & 26\% & 4\% \\
\hline
\end{tabular}
\end{table}

\subsection {Advantages of running at charm threshold}

The B factories, running at the $\Upsilon$(4S) will have produced 500 million
charm pairs by 2005. However, there are significant advantages of running at
charm threshold:
\begin{enumerate}
\item Charm events produced at threshold are extremely clean;
\item Double tag events, which are key to making absolute branching fraction
measurements, are pristine;
\item Signal/Background is optimum at threshold;
\item Neutrino reconstruction is clean;
\item Quantum coherence aids $D$ mixing and CP violation studies
\end{enumerate}
These advantages are dramatically illustrated in Figure~\ref{fig:cleoc_event},
which shows a picture of a simulated and fully reconstructed
$\psi(3770)\to D\bar{D}$ event.

\subsection {The CLEO-III Detector : Performance, Modifications and issues}

\begin{figure*}[t]
\epsfig{figure=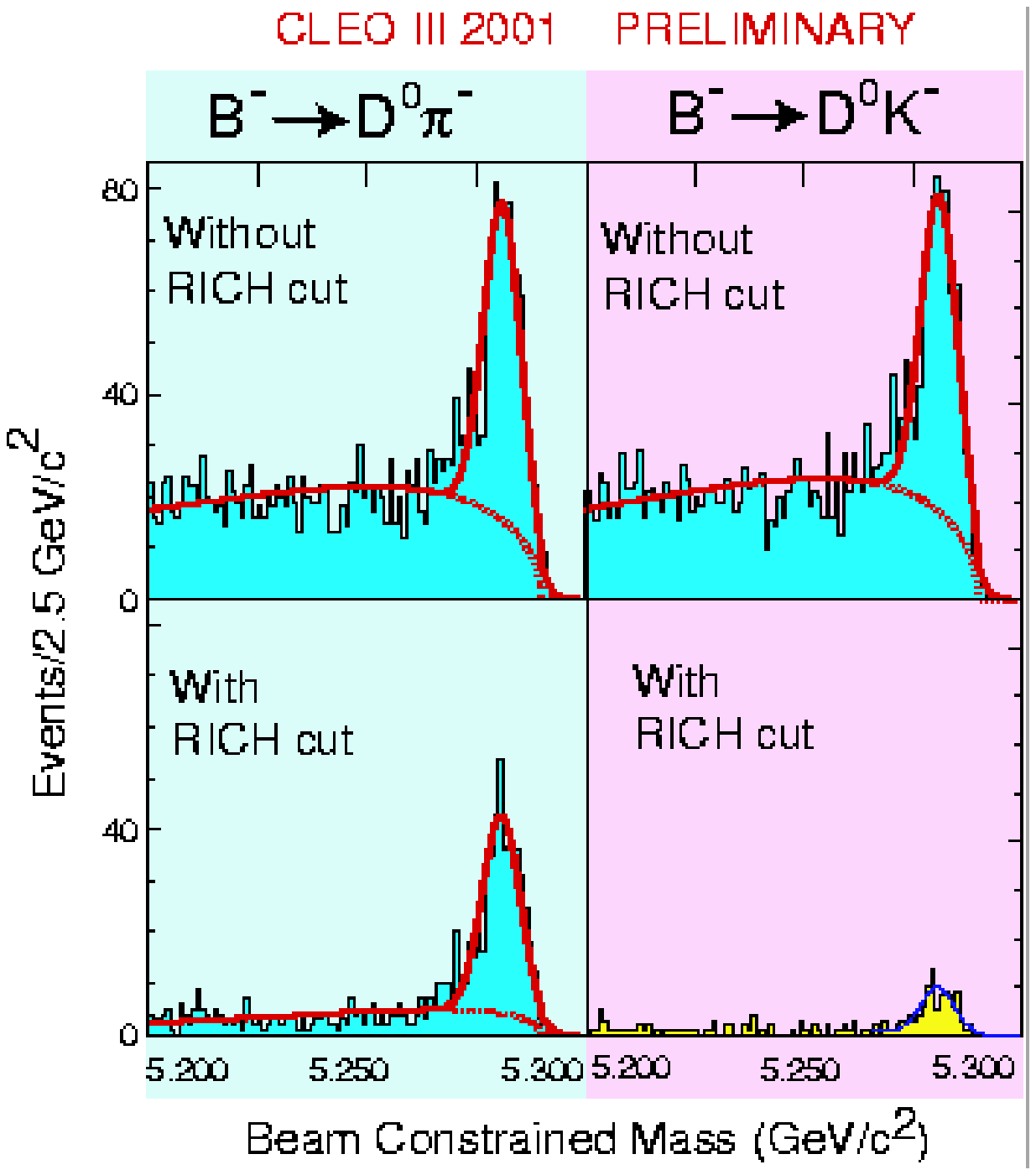,width=7.cm,height=7.cm}
\epsfig{figure=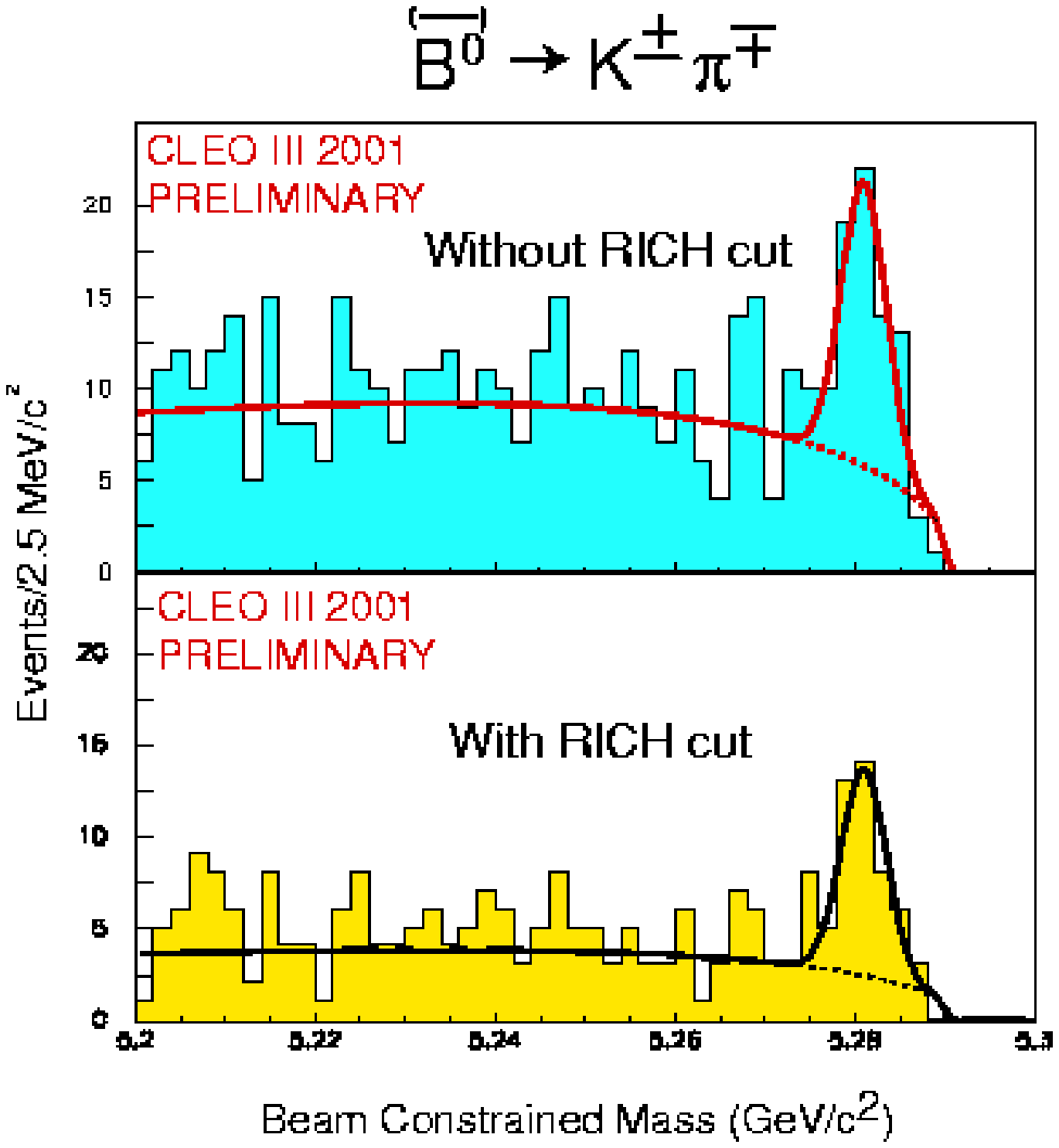,width=7.cm,height=7.cm} \caption
{(Left) Beam constrained mass for the Cabibbo allowed decay $B\to
D\pi$  and the Cabibbo suppressed decay $B\to DK$ with and without
RICH information. The latter decay was extremely difficult to
observe in CLEO II/II.V which did not have a RICH detector.
(Right) The penguin dominated decay $B \to K\pi$. Both of these
modes are observed in CLEO III with branching ratios consistent
with those found in CLEO II/II.V.} \label{fig:rich}
\end{figure*}

The CLEO III detector, shown in Figure~\ref{fig:cleo3_det}, consists of
a new silicon tracker, a new drift chamber,
and a Ring Imaging Cherenkov Counter (RICH), together  with the
CLEO II/II.V magnet, electromagnetic calorimeter and muon chambers.
The upgraded detector was installed and commissioned during the Fall of 1999
and Spring of 2000.
Subsequently operation has been very reliable (see below for a caveat)
and a very high quality data set has been obtained.
To give an idea of the power of the CLEO III detector
in Figure~\ref{fig:rich} (left plot) the beam
constrained mass for the Cabibbo allowed decay $B\to D\pi$ and
the Cabibbo suppressed decay $B\to DK$
with and without RICH information is shown.
The latter decay was extremely difficult to observe in
CLEO II/II.V which did not have a RICH detector.
In the right plot of Figure~\ref{fig:rich}
the penguin dominated decay $B \to K\pi$ and the tree dominated
decay is shown.
Both of these modes are observed in CLEO III with branching ratios
consistent with those found in CLEO
II/II.V. and are also in agreement with recent Belle and BABAR results.
Figure~\ref{fig:rich} is a demonstration that CLEO III performs very well indeed.

Unfortunately, there is one detector subsystem that is not performing well.
The CLEO III silicon has experienced an unexpected and unexplained loss of efficiency
The situation is under
constant evaluation. It is likely that the silicon detector
will be replaced with a wire vertex chamber for CLEO-c.
We note that if one was to design a charm factory detector from
scratch the tracking would be entirely gas based to ensure
that the detector material was kept to a minimum.
CLEO-c simulations indicate that a simple six layer stereo tracker
inserted into the CLEO III
drift chamber  as a silicon replacement would provide a system
with superior momentum resolution
to the current CLEO III tracking system.
We propose to build such
a device for CLEO-c.

Due to machine issues we plan to lower the solenoid field
strength to 1 T from 1.5 T. All other
parts of the detector do not require modification.
The dE/dx and Ring Imaging Cerenkov counters
are expected to work well over the CLEO-c momentum range.
The electromagnetic calorimeter
works well and has fewer photons to deal with at 3-5 GeV than at 10 GeV.
Triggers will work as before.
Minor upgrades may be required of the Data Acquisition system to
handle peak data transfer rates.
CESR conversion to CESR-c requires 18 m of wiggler magnets, to increase
transverse cooling,
at a cost of $\sim$ \$4M.
The conclusion is that, with the addition of the
replacement wire chamber, CLEO is expected to work well in the 3-5 GeV
energy  range at the expected rates.

\subsection {Examples of analyses with CLEO-c}

The main targets for the CKM physics program at CLEO-c are
absolute branching ratio
measurements of hadronic, leptonic and semileptonic decays.
The first of these provides an absolute
scale for all charm and hence all beauty decays.
The second measures decay constants and the third
measures form factors and in combination with theory allows
the determination of $V_{cd}$ and $V_{cs}$.

\subsubsection {Absolute branching ratios}

\begin{figure}
\epsfig{figure=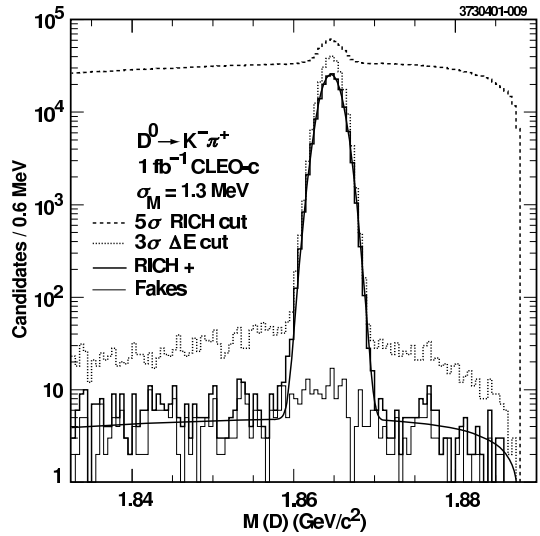,width=7cm,height=7cm}
\caption
{$K\pi$ invariant mass in $\psi(3770)\to D\bar{D}$ events
showing a strikingly clean signal for
$D\to K\pi$. The y axis is  logarithmic. The S/N $\sim$ 5000/1.}
\label{fig:d0kpi}
\end{figure}

\begin{figure}
\epsfig{figure=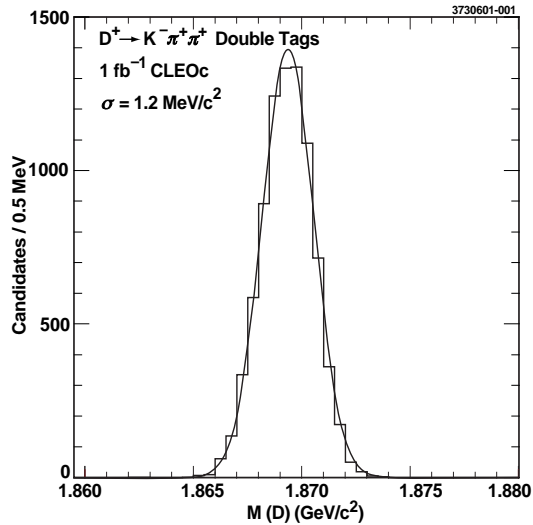,width=7cm,height=7cm}
\caption
{$K\pi\pi$ invariant mass in $\psi(3770)\to D\bar{D}$
events where the other D in the event has
already been reconstructed.
A clean signal for $D\to K\pi\pi$ is observed and the absolute
branching ratio $Br(D\to K\pi\pi)$ is measured by counting events
in the peak.}
\label{fig:br_dkpipi}
\end{figure}

The key idea is to reconstruct a $D$ meson in any hadronic mode.
This, then, constitutes the tag. Figure~\ref{fig:d0kpi}
shows tags in the mode $D\to K\pi$. Note the y axis is a log scale.
Tag modes are very clean. The signal to background ratio is $\sim$ 5000/1
for the example shown.
Since $\psi(3770) \to D\bar{D}$, reconstruction of a second D meson in a
tagged event to a final state X, corrected by the efficiency which is very
well known, and divided by
the number of D tags, also very well known, is a measure of
the absolute branching ratio
$Br(D\to X)$. Figure~\ref{fig:br_dkpipi} shows the $K^{-}\pi^{+}\pi^{+}$
signal from doubly tagged events. It is essentially background free.
The simplicity of $\psi(3770) \to D\bar{D}$
events combined with the absence of background allows the determination of
absolute branching ratios with extremely small systematic errors.
This is a key advantage of running at threshold.

\subsubsection {Leptonic decay $D_s\to\mu\nu$}

This is a crucial measurement because it provides information which
can be used to extract the weak decay constant, $f_{D_{s}}$. The
constraints provided by running at threshold are critical to extracting
the signal.

The analysis procedure is as follows:
\begin{enumerate}
\item Fully reconstruct one $D$;
\item Require one additional charged track and no additional photons;
\item Compute the missing mass squared (MM2)  which  peaks at zero for
a decay where only a neutrino is unobserved.
\end{enumerate}

\begin{figure*}
\epsfig{figure=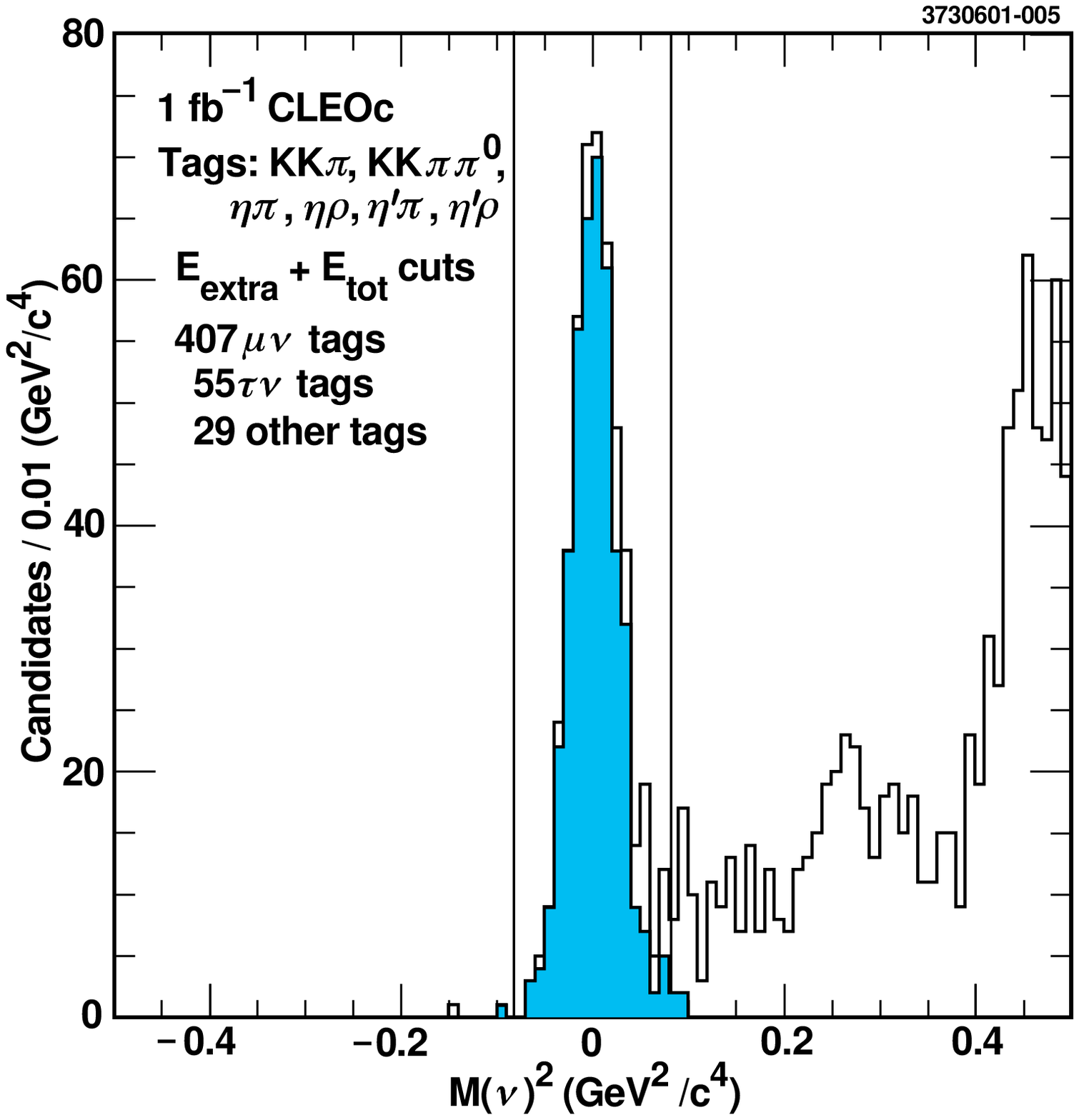,width=6cm,height=6cm}
\epsfig{figure=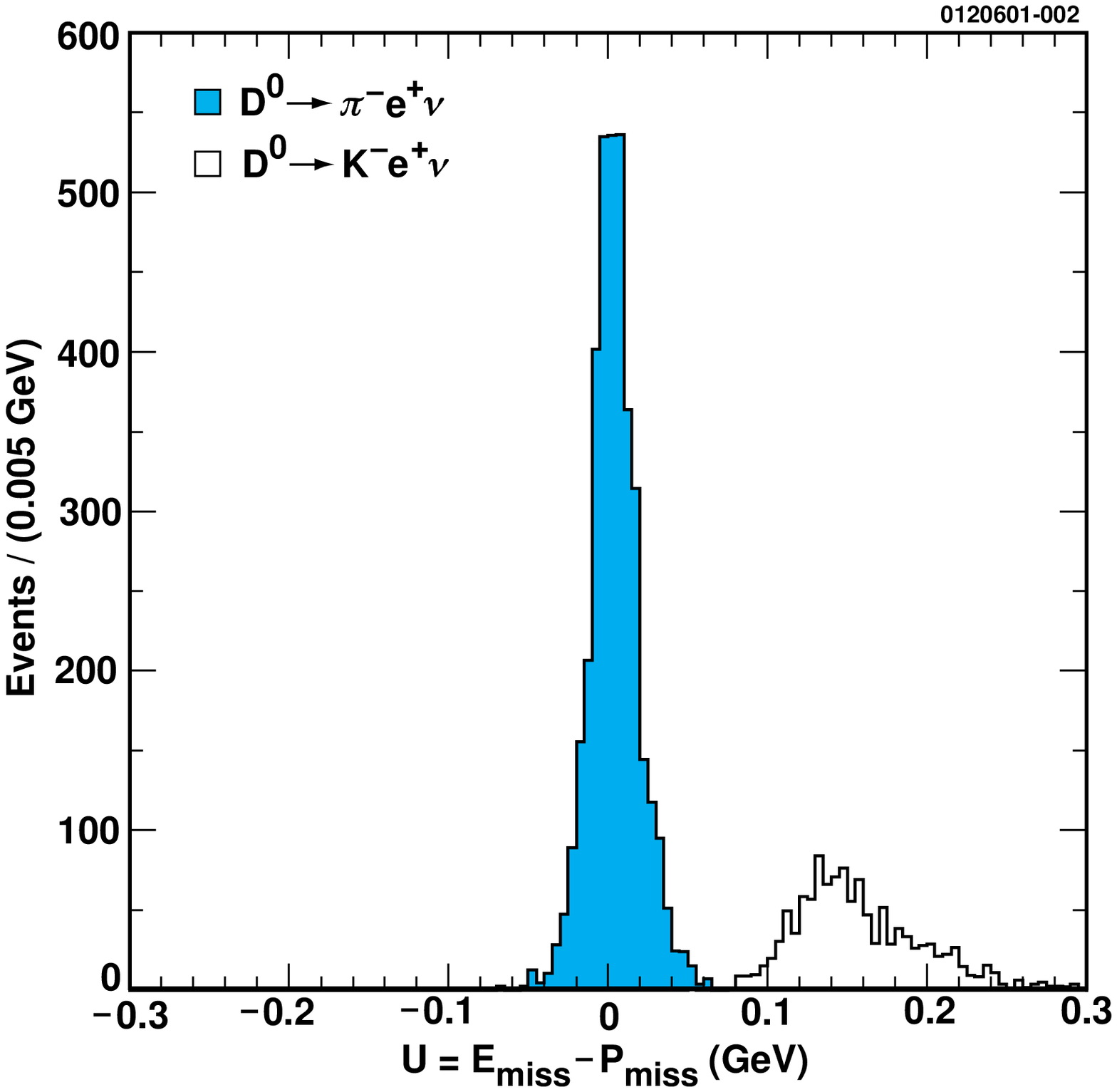,width=6cm,height=6cm}
\caption
{(Left) Missing mass for $D_{s}D_{s}$ tagged pairs produced
at $\sqrt{s}=4100$ MeV. Events due to the decay $D_s\to\mu\nu$ are shaded.
(Right) The difference between the missing energy and missing
momentum in $\psi(3770)$ tagged events for
the Cabibbo suppressed decay $D\to \pi \ell\nu$ (shaded).
The unshaded histogram arises from the ten times more copiously produced
Cabibbo allowed
transition $D\to K \ell\nu$ where the K is outside the fiducial volume of
the RICH.}
\label{fig:munu_pienu}
\end{figure*}

The missing mass resolution, which  is of order $\sim M_{\pi^0}$,
is sufficient to reject the backgrounds to this process as shown
in Fig.~\ref{fig:munu_pienu}. There is no need to identify muons,
which helps reduce the systematic error. One can inspect the
single prong to make sure it is not an electron. This provides a
check of the background level since the leptonic decay to an
electron is severely helicity-suppressed and no signal is expected
in this mode.

\subsubsection {Semileptonic decay $D\to \pi \ell\nu$}

The analysis procedure is as follows:
\begin{enumerate}
\item Fully reconstruct one D
\item Identify one electron and one hadronic track.
\item Calculate the variable, $U = E_{miss} - P_{miss}$, which
peaks at zero for semileptonic decays.
\end{enumerate}

Using the above procedure results in the right plot of
Figure~\ref{fig:munu_pienu}.
With CLEO-c for the first time it will become possible to make
absolute branching ratio and
absolute form factor measurements of every charm meson
semileptonic pseudoscalar to
pseudoscalar  and pseudoscalar to vector transition.
This will be a lattice calibration data set without
equal. Figure~\ref{fig:error_br} graphically shows the improvement
in absolute semileptonic branching ratios that CLEO-c will make.

\subsection {Run Plan}

CLEO-c must run at various center of mass energies to
achieve its physics goals. The ``run plan'' currently used
to calculate the physics reach is given below. Note that item 1 is prior
to machine conversion and the remaining items are post machine conversion.
\begin{enumerate}
\item 2002 : $\Upsilon$'s --  1-2 $fb^{-1}$ each at
$\Upsilon(1S),\Upsilon(2S),\Upsilon(3S)$ \\
Spectroscopy, electromagnetic transition matrix elements, the leptonic width.
$\Gamma_{ee}$, and searches for the yet to be discovered
$h_b, \eta_b$ with 10-20 times the existing world's data sample.
\item 2003 : $\psi(3770)$ -- 3 $fb^{-1}$ \\
30 million events, 6 million tagged D decays (310 times MARK III)
\item 2004 : 4100 MeV -- 3 $fb^{-1}$ \\
1.5 million $D_{s}D_{s}$ events, 0.3 million tagged $D_s$ decays
(480 times MARK III, 130 times BES)
\item 2005 : $J/\psi$ -- 1 $fb^{-1}$ \\
1 Billion $J/\psi$ decays (170 times MARK III, 20 times BES II)
\end{enumerate}

\subsection{Physics Reach of CLEO-c}

\begin{figure}
\epsfig{figure=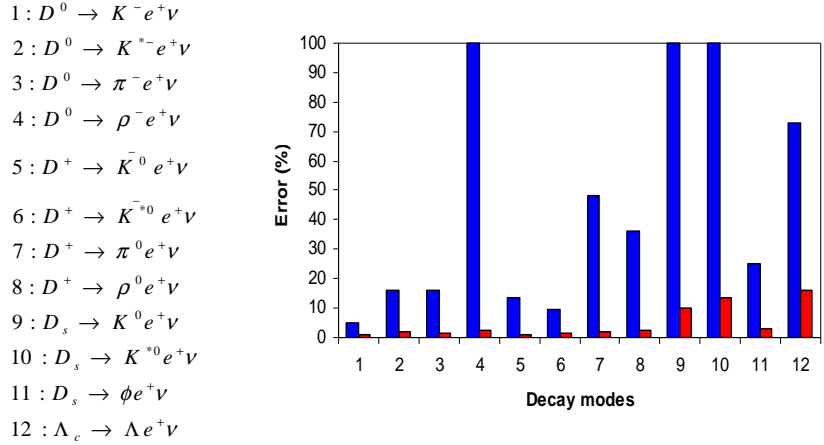,height=11cm,angle=-90}
\caption
{Absolute branching ratio current precision from the PDG
(left entry) and precision
attainable at CLEO-c (right entry ) for twelve semileptonic charm decays.}
\label{fig:error_br}
\end{figure}

\begin{figure}
\epsfig{figure=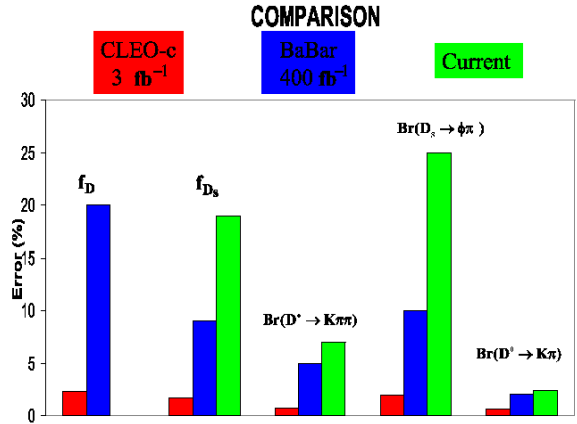,width=8cm,height=6cm}
\caption
{Comparison of CLEO-c  (left) BaBar (center) and PDG2001 (right)
for five physics quantities indiocated in the key.}
\label{fig:comparison}
\end{figure}

Tables~\ref{tab:charm}, \ref{tab:ckm} , and \ref{tab:comparison},
and Figures~\ref{fig:error_br} and
\ref{fig:comparison} summarize the CLEO-c measurements
of charm weak
decays, and compare the precision obtainable with CLEO-c
to the expected precision at BABAR
which expects to have recorded 500 million charm pairs by 2005.
CLEO-c clearly achieves far greater
precision for many measurements.
The reason for this is the ability to measure absolute branching
ratios by tagging and the absence of background at threshold.
In those analyses where CLEO-c is
not dominant it remains comparable or complementary to the B factories.

Also shown in Table~\ref{tab:comparison} is a summary of the data set size
for CLEO-c and BES II at the $J/\psi$ and $\psi'$,
and the precision with which R, the ratio of the $e^{+}e^{-}$ annihilation
cross section into hadrons to mu pairs, can be measured.
The CLEO-c data sets are over an order of magnitude larger, the precision
with which R is measured is a factor of three higher,
in addition the CLEO detector is vastly
superior to the BES II detector.
Taken together the CLEO-c datasets at the $J/\psi$ and $\psi'$ will be
qualitatively and quantitatively superior to any previous dataset
in the charmonium sector thereby
providing discovery potential for glueballs and exotics without equal.

\begin{table}
\caption{Summary of direct CKM reach with CLEO-c}
\label{tab:ckm}
\begin{tabular}{cccccc}
\hline
Topic & Reaction & Energy & $L $        & current     & CLEO-c \\
      &          & (MeV)  & $(fb^{-1})$ & sensitivity & sensitivity \\
\hline
$V_{cs}$ & $D^0\to K\ell^+\nu$ & 3770 & 3 & 16\% & 1.6\% \\
\hline
$V_{cd}$ & $D^0\to\pi\ell^+\nu$ & 3770 & 3 & 7\% & 1.7\% \\
\hline
\end{tabular}
\end{table}

\begin{table}
\caption{Comparision of CLEO-c reach to BaBar and BES}
\label{tab:comparison}
\begin{tabular}{cccccc}
\hline
Quantity & CLEO-c & BaBar  & Quantity & CLEO-c & BES-II \\
\hline
$f_D$ & 2.3\% & 10-20\%    & \#$J/\psi$ & $10^9$ & $5\times 10^7$ \\
\hline
$f_{D_s}$ & 1.7\% & 5-10\% & $\psi'$   & $10^8$ & $3.9\times 10^6$ \\
\hline
$Br(D^0 \to K\pi)$ & 0.7\% & 2-3\% & 4.14 GeV & $1 fb^{-1}$ &$23 pb^{-1}$\\
\hline
$Br(D^+ \to K\pi\pi)$ & 1.9\% & 3-5\%  & 3-5 R Scan & 2\%  & 6.6\% \\
\hline
$Br(D_s^+\to\phi\pi)$ & 1.3\% & 5-10\% & \multicolumn{3}{c}{} \\
\hline
\end{tabular}
\end{table}

\subsection{CLEO-c Physics Impact}

CLEO-c will provide crucial validation of Lattice QCD,
which  will be able to  calculate with
accuracies of 1-2\%.
The CLEO-c decay constant and semileptonic data will provide a ``golden'',
and timely test
while CLEO-c QCD and charmonium data provide additional benchmarks.

CLEO-c will provide dramatically improved  knowledge of absolute
charm branching fractions
which are now contributing significant errors to measurements involving
b's in a timely fashion.
CLEO-c will significantly improve knowledge of CKM matrix elements
which are now not very well known.
$V_{cd}$ and $V_{cs}$ will be determined directly by CLEO-c data and LQCD,
or other theoretical techniques.
$V_{cb}, V_{ub}, V_{td}$
and $V_{ts}$ will be determined with enormously improved precision
using B factory data once the
CLEO-c program of lattice validation is complete.
Table~\ref{tab:ckm_summary} give a summary of the situation.

CLEO-c data alone will also allow new tests of the unitarity of the CKM matrix.
The unitarity of the second
row of the CKM matrix will be probed at the 3\% level which
is comparable to our current knowledge of the first row.
CLEO-c data will also test unitarity by measuring
the ratio of the long sides of the
squashed $cu$ triangle to 1.3\%.

Finally the potential to observe new forms of matter; glueballs,
hybrids, etc in $J/\psi$ decays and new
physics through sensitivity to charm mixing, CP violation,
and rare decays provides a discovery
component to the program.

\begin{table}
\caption{Current knowledge of CKM matrix elements (row one). Knowledge of
CKM matrix elements after CLEO-c (row two).
See the text for further details.}
\label{tab:ckm_summary}
\begin{tabular}{cccccc}
\hline
$V_{cd}$ & $V_{cs}$ & $V_{cb}$ & $V_{ub}$ & $V_{td}$ & $V_{ts}$ \\
\hline
7\% & 16\% & 5\% & 25\% & 36\% & 39\% \\
\hline
1.7\% & 1.6\% & 3\% & 5\% & 5\% & 5\% \\
\hline
\end{tabular}
\end{table}


\begin{theacknowledgments}

I would like to thank Dave Hitlin and Frank Porter
for the superb organization of this conference.

\end{theacknowledgments}


          {\ifthenelse{\equal{\AIPcitestyleselect}{num}}
             {\bibliographystyle{arlonum}}
             {\bibliographystyle{arlobib}}
          }

\begin{thebibliography}{99}

\bibitem{cleo-c}
``CLEO-c and CESR-c : A New Frontier of Weak and Strong Interactions'',
CLNS 01/1742.

\bibitem{shipsey2}
``E2 Snowmass Report'', I. Shipsey on behalf of the E2 convenors,
G. Burdman, J. Butler, I. Shipsey, and H. Yamamoto. Talk at the
final plenary session of Snowmass, 2001, ``The Future of Particle
Physics'', Snowmass, CO, July 2001. All E2 working group talks
(refs. 2-11) may be found at
http://www.physics.purdue.edu/Snowmass2001\_E2/

\bibitem{artuso1}
``Another look at Charm: the CLEO-c physics program'',
M. Artuso, talk to the E2 Working Group.

\bibitem{shipsey}
``CLEO-c and CESR-c : A New Frontier of Weak and Strong Interactions'',
I. Shipsey, talk to a joint E2/P2/P5 Working Group session.

\bibitem{dytman}
``Projected Non-perturbative QCD Studies with CLEO-c'', S. Dytman, talk to
the E4 Working Group.

\bibitem{gibbons1}
``CLEO-c and R measurements'', L. Gibbons,
talk to the E2 Working Group.

\bibitem{gibbons2}
``An Introduction to CLEO-c '', L. Gibbons,
talk to the E2 Working Group.

\bibitem{cassel}
``CLEO-C reach in D meson Decays :
Measuring absolute D meson branching fractions,
D decay constants,and CKM matrix elements'', D. Cassel,
talk to the E2 Working Group.

\bibitem{artuso2}
``Beyond the Standard Model: the clue from charm'', M. Artuso,
talk to the E2 Working Group.

\bibitem{pordes}
``A case for running CLEO-C at the $\psi^{\prime}$ $(\sqrt(s) = 3686$ MeV)'',
S. Pordes, talk to the E2 Working Group.

\bibitem{maravin}
``Experimental Aspects of Tau Physics at CLEO-c'', Y. Maravin,
talk to the E2 Working Group.

\end{thebibliography}

\end{document}